\documentclass[prc,aps,twocolumn,showpacs,amssymb,superscriptaddress,fleqn]{revtex4-1}

\usepackage{amsmath}
\usepackage{color}
\usepackage{graphicx}
\usepackage{dcolumn}
\usepackage{mathtools}
\usepackage{relsize}
\usepackage{float}
\usepackage{braket}
\newcommand{\heff}{$H_{\rm eff}$}
\newcommand{\heffs}{$H_{\rm eff}$s}
\newcommand{\qbox}{$\hat{Q}$-box}
\newcommand{\vlwk}{$V_{{\rm low}\mbox{-}k}$}
\newcommand{\nmax}{$N_{\rm max}$}

\DeclareSymbolFont{largesymbol}{OMX}{yhex}{m}{n}
\DeclareMathAccent{\Widehat}{\mathord}{largesymbol}{"62}

\begin{document}

\title{A shell-model study of calcium isotopes towards their drip line}

\author{L. Coraggio}
\affiliation{Istituto Nazionale di Fisica Nucleare, \\
Complesso Universitario di Monte  S. Angelo, Via Cintia, I-80126 Napoli, Italy}
\author{G. De Gregorio}
\affiliation{Dipartimento di Matematica e Fisica, Universit\`a degli
  Studi della Campania ``Luigi Vanvitelli'', viale Abramo Lincoln 5 -
  I-81100 Caserta, Italy}
\affiliation{Istituto Nazionale di Fisica Nucleare, \\
Complesso Universitario di Monte  S. Angelo, Via Cintia, I-80126 Napoli, Italy}
\author{A. Gargano}
\affiliation{Istituto Nazionale di Fisica Nucleare, \\
Complesso Universitario di Monte  S. Angelo, Via Cintia, I-80126 Napoli, Italy}
\author{N. Itaco}
\affiliation{Dipartimento di Matematica e Fisica, Universit\`a degli
  Studi della Campania ``Luigi Vanvitelli'', viale Abramo Lincoln 5 -
  I-81100 Caserta, Italy}
\affiliation{Istituto Nazionale di Fisica Nucleare, \\ 
Complesso Universitario di Monte  S. Angelo, Via Cintia, I-80126 Napoli, Italy}
\author{T. Fukui}
\affiliation{Yukawa Institute for Theoretical Physics, Kyoto
  University,\\
  Kitashirakawa Oiwake-Cho, Kyoto 606-8502, Japan}
\author{Y. Z. Ma}
\affiliation{School of Physics, and State Key Laboratory of Nuclear
  Physics and Technology, \\
Peking University, Beijing 100871, China}
\author{F. R. Xu}
\affiliation{School of Physics, and State Key Laboratory of Nuclear
  Physics and Technology, \\
Peking University, Beijing 100871, China}
 
\begin{abstract}
We report in this paper a study in terms of the nuclear shell model
about the location of the calcium isotopes drip line.
The starting point is considering the realistic two-body potential
derived by Entem and Machleidt within chiral perturbation theory at
next-to-next-to-next-to-leading order (N$^3$LO), as well as a chiral
three-body force at next-to-next-to-leading order (N$^2$LO) whose
structure and low-energy constants are consistent with the two-body
potential.
Then we construct the effective single-particle energies and residual
interaction needed to diagonalize the shell-model Hamiltonian.
The calculated two-neutron separation energies agree nicely with
experiment until $^{56}$Ca, which is the heaviest isotope whose mass
has been measured, and do not show any sign of two-neutron emission
until $^{70}$Ca.
We discuss the role of the choice of the model space in determining
the neutron drip line, and also the dependence of the results on the
parameters of the shell-model Hamiltonian.
\end{abstract}

\pacs{21.60.Cs, 21.30.Fe, 21.45.Ff, 27.40.+z}

\maketitle

\section{Introduction}
\label{intro}
The location of the boundaries of the nuclear landscape is an important
target in order to understand the limits of the strong force to hold
together the nucleons in a bound system.

The search of drip lines is not an easy task from the experimental
point of view, due to the extremely low production rates in
investigations dealing with the fragmentation of stable nuclei, and
the subsequent separation and identification of the products.
This is why the possibility of expanding the chart of nuclides is
closely linked to the development of facilities providing a new
generation of   re-accelerated radioactive ion beams as well as of new
instrumentation and advanced techniques (see, for example,
Ref. \cite{Thoennesen13}).
Consequently, theory is in charge to provide further insight on this
topic, and also to look for useful tips for experimental studies whose
results may help to find evidence of proton/neutron drip lines. 

In this context, calcium isotopes are one of the most intriguing
subjects to be investigated, since this isotopic chain spans from the
neutron-deficient $^{36}$Ca, which lies close to the proton drip line,
up to $^{60}$Ca, that has been recently observed \cite{Tarasov18} and
whose $N/Z$ ratio, equal to 2, classifies it as a very exotic nuclear
system.
As a matter of fact, the observation of $^{60}$Ca draws a line
between studies predicting it as a loosely bound nucleus
\cite{Meng02,Hagen12,Hergert14}
and those pushing neutron drip line up to $^{70}$Ca
\cite{Bhattacharya05,Chen15,Cao19,Neufcourt19}.
Moreover, recent mass measurements of heavy calcium isotopes
\cite{Wienholtz13,Michimasa18} are helpful to narrow the spread of the
theoretical predictions about this yet undiscovered section of the
chart of nuclides.

Another interesting aspect is the impulse that the novel experimental
results have given to the advance in theory.
In Ref. \cite{Neufcourt19} the authors have followed an innovative
approach to the study of calcium-isotopes drip line; they have
performed a model averaging analysis of the outcome of different
density-functional calculations, by way of Bayesian machine learning.
The result of their study is that, by considering both experimental
information and current density-functional models, 
$^{68}$Ca owns an average posterior probability of about 76$\%$ to be
bound with respect to the two-neutron emission.
It is worth stressing that this study conjugates advances in
computational theory - machine-learning methods - with a comparative
analysis of energy-density functionals that are constrained to
reproduce a variety of nuclear binding energies and radii.

Nuclei approaching the neutron drip line may show exotic features such
as an extended neutron distribution and halo.
This may be difficult to be reproduced with the harmonic-oscillator
basis, since this could provide a slow convergence rate of the calculations.
On the above grounds, the authors in Ref. \cite{Hagen12} have
investigated the evolution of shell structure of neutron-rich calcium
isotopes by way of the coupled-cluster method, starting from
chiral two- and three-body forces, and including the coupling to the
particle continuum in terms of the Berggren basis.

The effects of a neutron skin have been studied in
Ref. \cite{Hagen13}, where the neutron $^{60}$Ca $S$-wave scattering
phase shifts have been calculated within the coupled-cluster theory,
employing interactions derived from chiral perturbation theory.
The authors have found evidence of Efimov physics, namely a discrete
scale invariance in three-body systems such as a tight core and two
loosely bound nucleons \cite{Efimov70}.

Neutron-halo features of neutron-rich calcium isotopes have been also
investigated within the framework of the relativistic-mean-field and
complex-momentum-representation method \cite{Cao19}, providing
indications of a possible halo or giant halo structure of isotopes
with $N \geq 40$. 

Finally, it is also worth mentioning recent studies about the calcium
isotopic chain by way of microscopic many-body approaches - and
employing chiral two- and three-body potentials - such as the
Bogoliubov many-body perturbation theory \cite{Tichai18}, the
In-Medium Similarity Renormalization Group \cite{Simonis17,Hoppe19},
and the Self-Consistent Green's Function  Theory \cite{Soma20}.

Our aim is to study heavy calcium isotopes, providing a prediction of
their neutron drip line and shell evolution by way of nuclear shell
model (SM).
The framework is the same of our previous study about the
monopole component of the shell-model Hamiltonian for $0f1p$-shell
nuclei, where the single-particle (SP) energies and the two-body
matrix elements (TBMEs) of the residual interaction have been derived
from two- and three-body forces obtained by way of the chiral
perturbation theory (ChPT) \cite{Ma19}.
The main difference here is that we consider a larger model space, by
including the neutron $0g_{9/2}$ orbital in addition to the $0f1p$
orbitals, a choice that is needed to extend the calculations to
isotopes with $N > 40$.
We will focus on the relevance of the new SP degree of freedom, and of
the induced three-body contributions that appear in the effective SM
Hamiltonian \heff~to account for many-body correlations in nuclei with
more than two valence nucleons \cite{Polls83}.
This means that in the derivation of \heff we consider also the
interaction of clusters of three-valence nucleons with core
excitations as well as with virtual intermediate nucleons scattered
above the model space, through density-dependent \heffs, whose TBMEs
change according to the number of valence nucleons.

It should be pointed out that a similar approach has been followed by
Holt and coworkers in Refs. \cite{Holt12,Holt14}, where the two-body
chiral potential has been renormalized through the \vlwk~technique
\cite{Bogner02} and the effect of many-body correlations in the
derivation of \heff has been neglected.

This paper is organized as follows: next section is devoted to present
a few details of the derivation of the effective SM Hamiltonian from
realistic two- and three-body nuclear potentials, that is framed
within the many-body perturbation theory.
In Section \ref{results} the results of the diagonalization of the
\heffs~ for the calcium isotopic chain are presented and compared with
available data from experiment.
We will also show the results obtained previously within a smaller
model space \cite{Ma19}, those obtained neglecting many-body
correlations, and varying the energy of the $0g_{9/2}$ orbital, in
order to provide insight on the sensitivity of SM results to the
degrees of freedom our \heffs~ account for.
Finally, in Section \ref{endings} we draw the conclusions of present
investigation.

\section{Outline of calculations}\label{theory}
\subsection{The effective shell-model Hamiltonian}
The SM parameters, that are needed to diagonalize the SM Hamiltonian,
are derived from realistic nucleon-nucleon ($NN$) and three-body
($NNN$) potentials, both of them derived within the ChPT at
next-to-next-to-next-to-leading order (N$^3$LO) \cite{Entem02} and at
next-to-next-to-leading order (N$^2$LO), respectively.
These potentials consistently share the same nonlocal regulator
function, and some low-energy constants (LECs).
More precisely,  the N$^2$LO $NNN$ potential is composed of
three components, namely the two-pion ($2\pi$) exchange term
$V_{3NF}^{(2\pi)}$, the one-pion ($1\pi$) exchange plus contact term
$V_{3NF}^{(1\pi)}$, and the contact term $V_{3NF}^{\textrm{(ct)}}$.
The $c_1$, $c_3$, and $c_4$ LECs which characterize these terms
are the same as those in the N$^3$LO $NN$ potential, and are
determined by the renormalization procedure that is employed to fit
the $NN$ data \cite{Machleidt11}.

The values of the additional LECs appearing in 1$\pi$-exchange and
contact terms of the $NNN$ potential, $c_D$ and $c_E$, have been
chosen as $c_D=-1$ and $c_E=-0.34$.
They have been determined in no-core shell model calculations
\cite{Navratil07a,Maris13}, by identifying a set of observables in light
$p$-shell nuclei that are strongly sensitive to the $c_D$ value, and
then $c_E$ has been constrained to reproduce the binding energies of
the $A=3$ system.

In Appendix of Ref. \cite{Fukui18}, the details of the calculation of
matrix elements of the N$^2$LO $NNN$ potential, with a nonlocal
regulator, in a harmonic-oscillator (HO) basis can be found.

The Coulomb potential is explicitly taken into account aside the
matrix elements of the $NN$ potential.
The oscillator parameter $\hbar \omega$ we have employed to compute
the matrix elements of the $NN$ and $NNN$ potentials in the HO
oscillator basis is equal to 11 MeV, according to the expression
$\hbar \omega= 45 A^{-1/3} -25 A^{-2/3}$  for $A=40$
\cite{Blomqvist68}.

These nuclear potentials are the foundations to build up the effective
SM Hamiltonian \heff~that provides SP energies and TBMEs to solve the
SM eigenvalue problem.
As is well known, \heff~accounts for the degrees of freedom that are
not explicitly included in the truncated Hilbert space of the
configurations that, in our case, is spanned by four $0f1p$ plus
$0g_{9/2}$ neutron orbitals outside the doubly-closed $^{40}$Ca.

To this end, we need a similarity transformation which arranges,
within the full Hilbert space of the configurations, a decoupling of
the model space $P$ where the valence nucleons are constrained from
its complement $Q=1-P$.

We tackle this problem within the time-dependent perturbation theory,
namely \heff~is expressed through the Kuo-Lee-Ratcliff folded-diagram
expansion in terms of the \qbox~vertex function
\cite{Kuo90,Hjorth95,Coraggio12a}.

The \qbox~is defined in terms of the full nuclear Hamiltonian
$H=H_0+H_1$, where $H_0$ is the unperturbed component and $H_1$ the
interaction one:

\begin{equation}
\hat{Q} (\epsilon) = P H_1 P + P H_1 Q \frac{1}{\epsilon - Q H Q} Q
H_1 P ~, \label{qbox}
\end{equation}
\noindent
and $\epsilon$ is an energy parameter called ``starting energy''.

Since the exact calculation of the \qbox~is impossible, the term
$1/(\epsilon - Q H Q)$ is expanded as a power series

\begin{equation}
\frac{1}{\epsilon - Q H Q} = \sum_{n=0}^{\infty} \frac{1}{\epsilon -Q
  H_0 Q} \left( \frac{Q H_1 Q}{\epsilon -Q H_0 Q} \right)^{n} ~,
\end{equation}

\noindent
leading to the perturbative expansion of the \qbox.
It is useful to employ a diagrammatic representation of this
perturbative expansion, which is a collection of Goldstone diagrams
that have at least one $H_1$-vertex, are irreducible - namely at least
one line between two successive vertices does not belong to the model
space - and are linked to at least one external valence line (valence
linked) \citep{Kuo71}.

Then, the \qbox~is employed to solve non-linear matrix equations
to derive \heff~ by way of iterative techniques such as the
Kuo-Krenciglowa and Lee-Suzuki ones \cite{Suzuki80}, or
graphical non-iterative methods \cite{Suzuki11}.
We have experienced that the latter provide a faster and more stable
convergence to the solution of the matrix equation to derive \heff,
and are the ones we have employed in present work.

We include in our $\hat{Q}$-box expansion one- and two-body Goldstone
diagrams through third order in the $NN$ potential and up to first
order in the $NNN$ one.
A complete list of diagrams with $NN$ vertices can be found in
Ref. \cite{Coraggio12a}, while the diagrams at first order in $NNN$
potential, as well as their analytical expressions, are reported in
Refs. \cite{Fukui18,Ma19}.
It is worth pointing out that these expressions are the coefficients
of the one-body and two-body terms arising from the normal-ordering
decomposition of the three-body component of a many-body Hamiltonian
\cite{HjorthJensen17}.
In Ref. \cite{Holt14}, Holt and coworkers, using a similar approach to
study Ca isotopes, have shown that the uncertainty linked to
neglecting contributions beyond the normal-ordered two-body components
(residual $NNN$ forces) is small.

Since we are going to study many-valence nucleon systems, we should
derive many-body \heffs~which depend on the number of valence
particles.
This means that the \qbox~should include at least contributions from
three-body diagrams accounting for the interaction via the two-body
force of the valence nucleons with configurations outside the model
space.

\begin{figure}[h]
\begin{center}
\includegraphics[scale=0.5,angle=0]{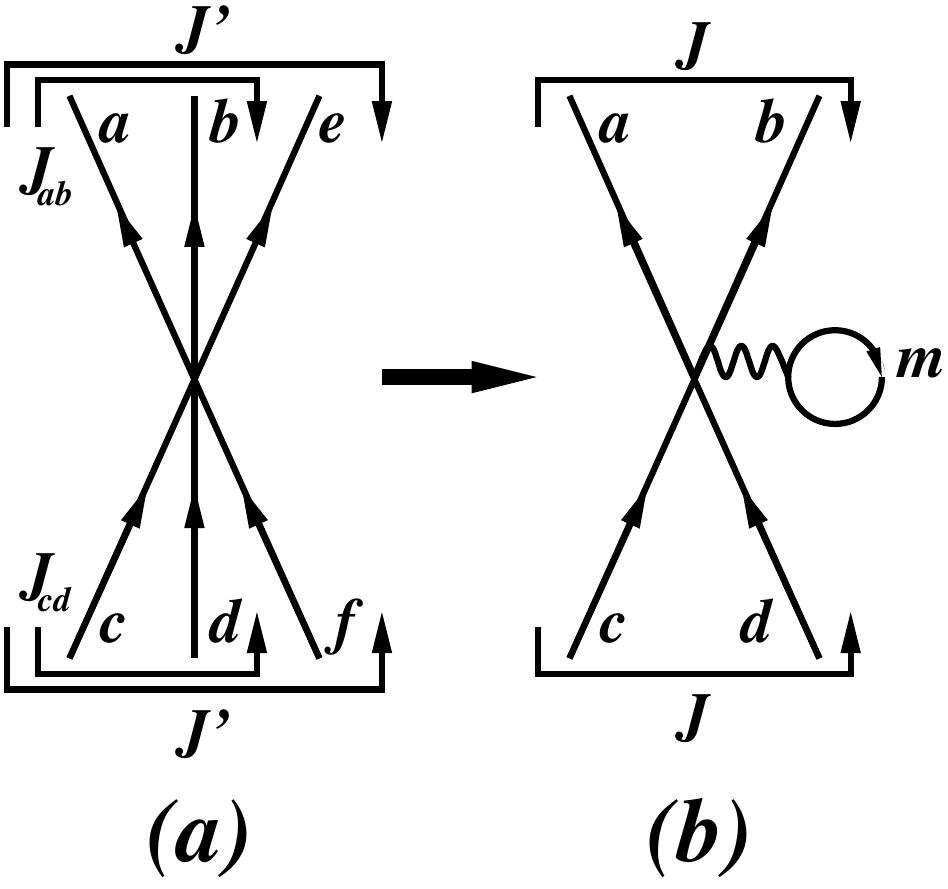}
\caption{Density-dependent two-body contribution that is
  obtained from a three-body one. Graph $(b)$ is obtained by
  summing over one incoming and outgoing particle of the three-body
  graph $(a)$.}
\label{3bf}
\end{center}
\end{figure}

Since we employ SM codes which cannot perform the diagonalization of a
three-body \heff~\cite{ANTOINE,KSHELL}, we derive a density-dependent
two-body term from the three-body contribution arising at second order
in perturbation theory.
Namely, nine one-loop diagrams (see the graph $(b)$ in
Fig. \ref{3bf}) are calculated from the corresponding
diagrams reported in Fig. 3 of Ref. \cite{Polls83}.

Their explicit form is reported in Ref. \cite{Ma19}, and depends on
the unperturbed occupation density $\rho_m$ of the orbital $j_m$,
leading to the derivation of \heffs~depending on the number of valence
protons and neutrons.
The density-dependent \heffs~differ only in their TBMEs since, as can
be seen in Fig. \ref{3bf}, these one-loop diagrams are two-body terms.

It should be recalled that, since the neutron $0g_{9/2}$ orbital may
couple with the $0f_{7/2}$ one to the total angular momentum
$J^{\pi}=1^-$, the results of the diagonalization of the shell-model
Hamiltonian within the model space with 5 neutron orbitals might be
affected somehow by the spurious center-of-mass motion
\cite{Elliott55}.

In order to check if these spuriosities are under control, we have
also performed calculations to separate in energy the excitations
originated by the internal degrees of freedom from those with spurious
center-of-mass components by following the procedure suggested by
Gloeckner and Lawson \cite{Gloeckner74}.

According to Ref. \cite{Gloeckner74}, the modified shell-model
Hamiltonian $H'$ should be diagonalized:

\begin{equation}
  H' = H_{\rm eff} + H_{\beta} ~~,
\label{hcom}
\end{equation}

\noindent
where $H_{\beta}$ is $\beta$ times the center-of-mass excitation
energy of the $A$-nucleon system

\begin{equation} 
H_{\beta} = \beta \left\{ \frac{(\sum_{i=1}^A {\bf p_i})^2 }{2Am} +
\frac{m \omega^2}{2A} (\sum_{i=1}^A {\bf r_i})^2 -\frac{3}{2} \hbar
\omega \right\}~~.
\end{equation}

The spurious components are then pushed up in energy by increasing the
parameter $\beta$, so that one can assume that the low-energy spectrum
is weakly influenced by the above components.
We have performed calculations using values of $\beta$ such that
$\gamma=\beta\frac{\hbar \omega}{A}$ is equal to 10 MeV and 15 MeV in
order to evaluate the role of the center-of-mass spuriosities \cite{Liu12}, and the
results will be reported in Section \ref{results}.

\subsection{Convergence properties of \heff}
We now discuss the convergence properties of our \heffs, an issue that
needs to be examined since the input chiral $NN$ and $NNN$ potentials
have not been modified by way of any renormalization procedure.

In Ref. \cite{Ma19} we have extensively discussed the order-by-order
behavior of the perturbative expansion of the \qbox, as well as the
convergence with respect to the dimension of the space of the
intermediate states, considering the systems with one- and two-valence
neutrons, namely $^{41}$Ca and $^{42}$Ca, respectively.
We recall that we express the number of intermediate states as a
function of the maximum allowed excitation energy of the intermediate
states expressed in terms of the oscillator quanta
\nmax~\cite{Coraggio12a}, and include intermediate states with an
unperturbed excitation energy up to $E_{\rm max}=N_{\rm max} \hbar
\omega$.
Because of our present limitation of the storage of the total number
of TBMEs, we can include, for $^{40}$Ca core, a maximum number of
intermediate states that do not exceed \nmax=18.

In Ref. \cite{Ma19}, we have shown that this value is not large enough 
to provide convergence of the SP spectrum of $^{41}$Ca, since the
chosen HO parameter is 11 MeV and the cutoff of both $NN$ and $NNN$
potentials is slightly larger than 2.5 fm$^{-1}$, this values leading
to a value of \nmax~to be at least 26.
However, this does not affect the convergence of the energy spacings
that are stable with respect to the increase in the number of
intermediate states from $N_{\rm max} \approx 12 - 14$ on for both
$^{41}$Ca and $^{42}$Ca, as already shown in Ref. \cite{Ma19}.

In this work we will show the convergence properties of \heff~by
considering a more complex system, such as $^{50}$Ca, which is
characterized by 10 valence neutrons.
By studying this system, we can test the perturbative behavior of
\heffs~that include density-dependent contributions which account for
three-body correlations.
It is worth pointing out that our choice is challenging, since
$^{50}$Ca exhibits a structure that is more collective than other neighbor
isotopes, whose closure properties provide a simpler structure of the
wave functions.

\begin{figure}[h]
\begin{center}
\includegraphics[scale=0.35,angle=0]{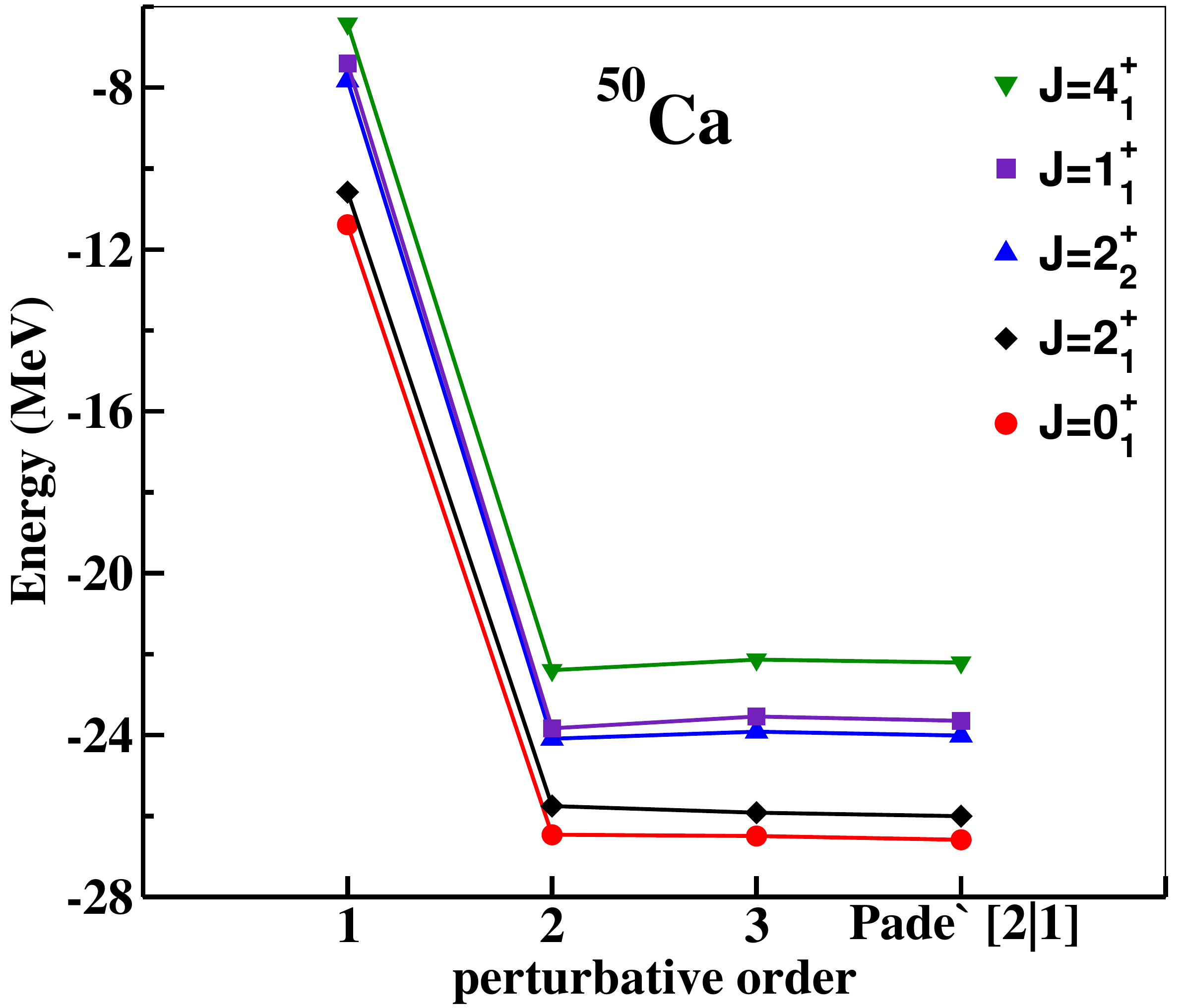}
\caption{Low-lying energy spectrum of $^{50}$Ca, obtained starting
  from $\hat{Q}$-boxes at first-, second-, and third-order in
  perturbation theory, and their Pad\'e approximant $[2|1]$.}
\label{50Ca_obo}
\end{center}
\end{figure}

In Fig. \ref{50Ca_obo} we report the low-lying states of $^{50}$Ca
spectrum, which have been obtained employing \heffs~starting from
$\hat{Q}$-boxes at first-, second-, and third-order in perturbation
theory, and their Pad\'e approximant $[2|1]$ \cite{Baker70}, while the
number of intermediate states is the largest we can manage,
i.e. \nmax=18.

We employ the Pad\'e approximant in order to obtain a better estimate
of the convergence value of the perturbation series
\cite{Coraggio12a}, as suggested in Ref. \cite{Hoffmann76}.

The results show a very satisfactory perturbative behavior of
\heff~with respect to the order-by-order convergence.

We consider now the dependence of \heff~as a function of the number of
intermediate states included in the calculation of the \qbox~second-
and third-order diagrams.

As it as been mentioned before, \nmax=18 does not provide a convergent
SP spectrum of $^{41}$Ca, but in Ref. \cite{Ma19} we have shown that
the SP spacings are stable.
Consequently, from now on for our calculations we consider 
SP spacings obtained from the theory but the value of the SP energy
of the neutron $0f_{7/2}$ orbital is fixed at -8.4 MeV, consistently
with the experimental value in $^{41}$Ca \cite{Audi03}.

\begin{figure}[h]
\begin{center}
\includegraphics[scale=0.30,angle=0]{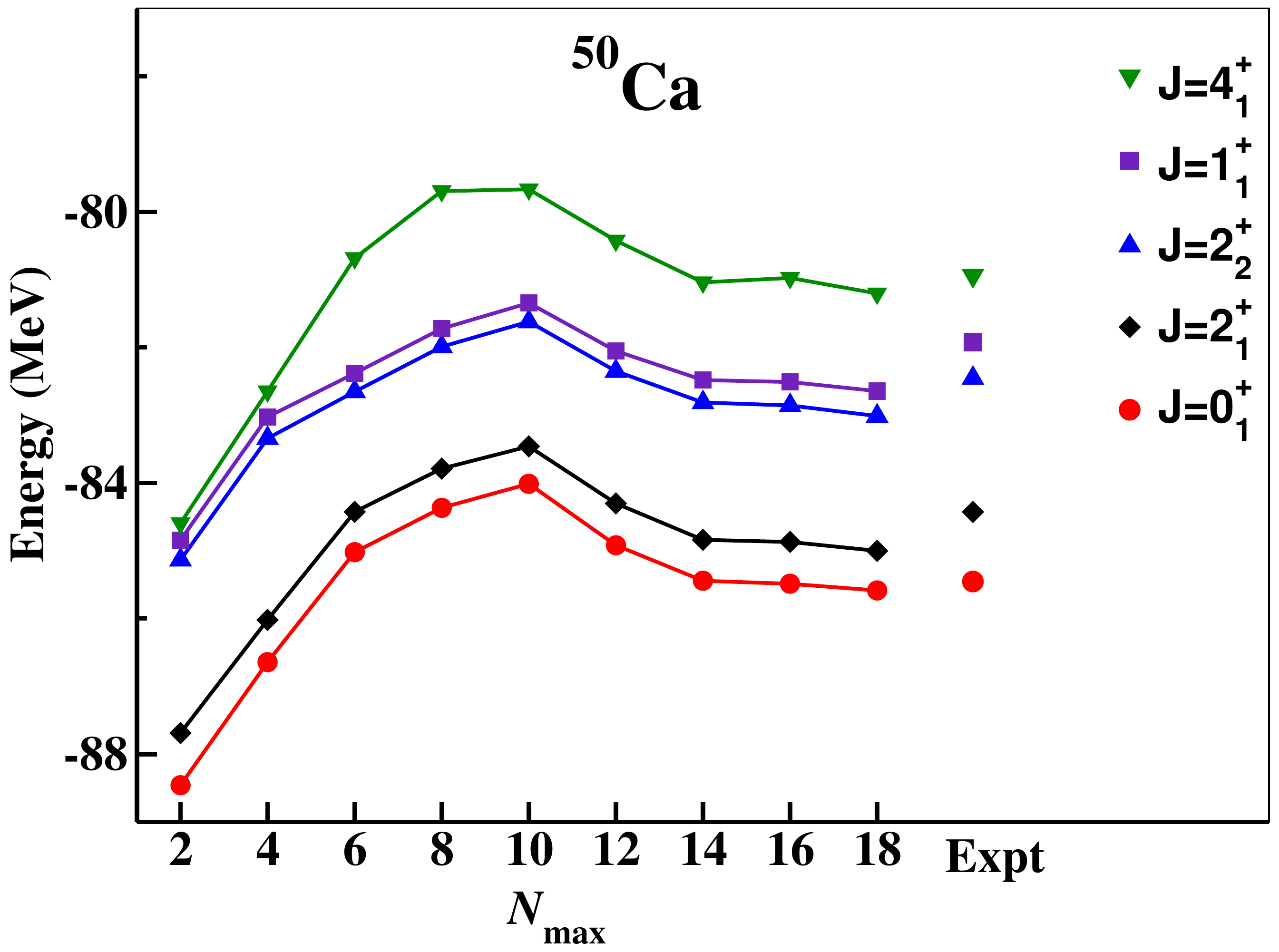}
\caption{Low-lying energy spectrum of $^{50}$Ca as a function of the
  number of intermediate states included in the perturbative
  calculation of the $\hat{Q}$-box.}
\label{50Ca_Nmax}
\end{center}
\end{figure}

In Fig. \ref{50Ca_Nmax} they are reported the energy spectra of
$^{50}$Ca, obtained diagonalizing the density-dependent
\heff~calculated employing the Pad\'e approximant $[2|1]$ of the
$\hat{Q}$-box, and including a number of intermediates states ranging
from \nmax=2 to 18.
Theoretical results are also compared with experiment \cite {ensdf}.

\begin{table}[ht]
  \caption{Theoretical proton ($\epsilon_{\pi}$) and neutron
    ($\epsilon_{\nu}$) SP energies (in MeV), referred to the
    $0f_{7/2}$ orbital.}
\begin{ruledtabular}
  \begin{tabular}{cccc}
 ~ & $\epsilon_{\pi}$ & ~ & $\epsilon_{\nu}$ \\
    \colrule
  ~ & ~ & ~ & ~ \\
 $0f_{7/2}$   & 0.0 & ~ & 0.0 \\ 
 $0f_{5/2}$   & 6.3 & ~ & 8.2 \\ 
 $1p_{3/2}$   & 2.5 & ~ & 3.2 \\ 
 $1p_{1/2}$   & 4.4 & ~ & 5.3 \\ 
 $0g_{9/2}$   & ~ & ~ & 9.9 \\ 
\end{tabular}
\end{ruledtabular}
\label{tablespe}
\end{table}

This is a test for our theoretical SP spacings and, especially, TBMEs,
and we observe that $^{50}$Ca spectrum converges from \nmax=14 on.
This leads to the conclusion that the SP spacings and TBMEs of our
\heff, calculated with \nmax=18, can be considered substantially
stable.
It is worth pointing out that in Ref. \cite{Ma19} the same study has
been performed for two-valence neutron system $^{42}$Ca, and in that
case the convergence rate is much faster.

We conclude this section by reporting in Table \ref{tablespe} the
proton and neutron SP energies $\epsilon_{\pi},\epsilon_{\nu}$,
calculated with respect to $0f_{7/2}$.
The proton-proton and proton-neutron \heff~channel has been derived by
considering a proton model space composed by the four orbitals
belonging to the $0f1p$ shell.
In the Supplemental Material \cite{supplemental2020} the TBMEs of
\heffs~ for systems with 2 and 10 valence nucleons can be found.

\section{Results}
\label{results}
In our previous study about nuclei belonging to $0f1p$ shell, we have
evidenced the crucial role played by the $NNN$ component of nuclear
Hamiltonians derived by way of ChPT, in order to provide SP energies
and TBMEs that may reproduce the shell evolution as observed from the
experiment \cite{Ma19}.
We have seen that SP energies and TBMEs of \heff~derived only from
the $NN$ component own deficient monopole components, which cannot
provide the shell closures at $N=28$ for both $^{48}$Ca and $^{56}$Ni.

\begin{table*}[ht]
\caption{Calculated and experimental two-neutron separation energies (in
  MeV). Data are taken from Refs. \cite{Audi03,Wienholtz13,Michimasa18}.}
\begin{ruledtabular}
  \begin{tabular}{lccccccccccccccc}
 ~ & $^{42}$Ca & $^{44}$Ca & $^{46}$Ca & $^{48}$Ca & $^{50}$Ca &
$^{52}$Ca & $^{54}$Ca & $^{56}$Ca & $^{58}$Ca & $^{60}$Ca &
$^{62}$Ca & $^{64}$Ca & $^{66}$Ca & $^{68}$Ca & $^{70}$Ca \\
    \colrule
Calculated $S_{2n}$ & 18.562 & 18.514 & 18.351 & 18.105 &
 12.054 & 11.754 & 6.914 & 3.377 & 3.347 & 3.017 & 2.266 &
   2.202 &  2.526  & 3.048 & 3.605 \\ 
Experimental $S_{2n}$   & 19.844 & 20.064 & 17.810 & 17.218 &
 11.517 & 10.73 & 7.127 & 4.492 & ~ & ~ & ~ & ~ &  ~  & ~ & ~ \\ 
\end{tabular}
\end{ruledtabular}
\label{tableS2n}
\end{table*}

On the above grounds, in our present work we are going to deal only
with \heffs~that are derived with $NN$ and $NNN$ components.

We start showing in Table \ref{tableS2n} our calculated two-neutron
separation energies up to $^{70}$Ca, compared with the available
experimental data \cite{Audi03,Wienholtz13,Michimasa18}.
As previously mentioned, the neutron SP energies reported in Table
\ref{tablespe} are shifted to reproduce the experimental g.s. energy
of $^{41}$Ca with respect to $^{40}$Ca.

\begin{figure}[H]
\begin{center}
\includegraphics[scale=0.31,angle=0]{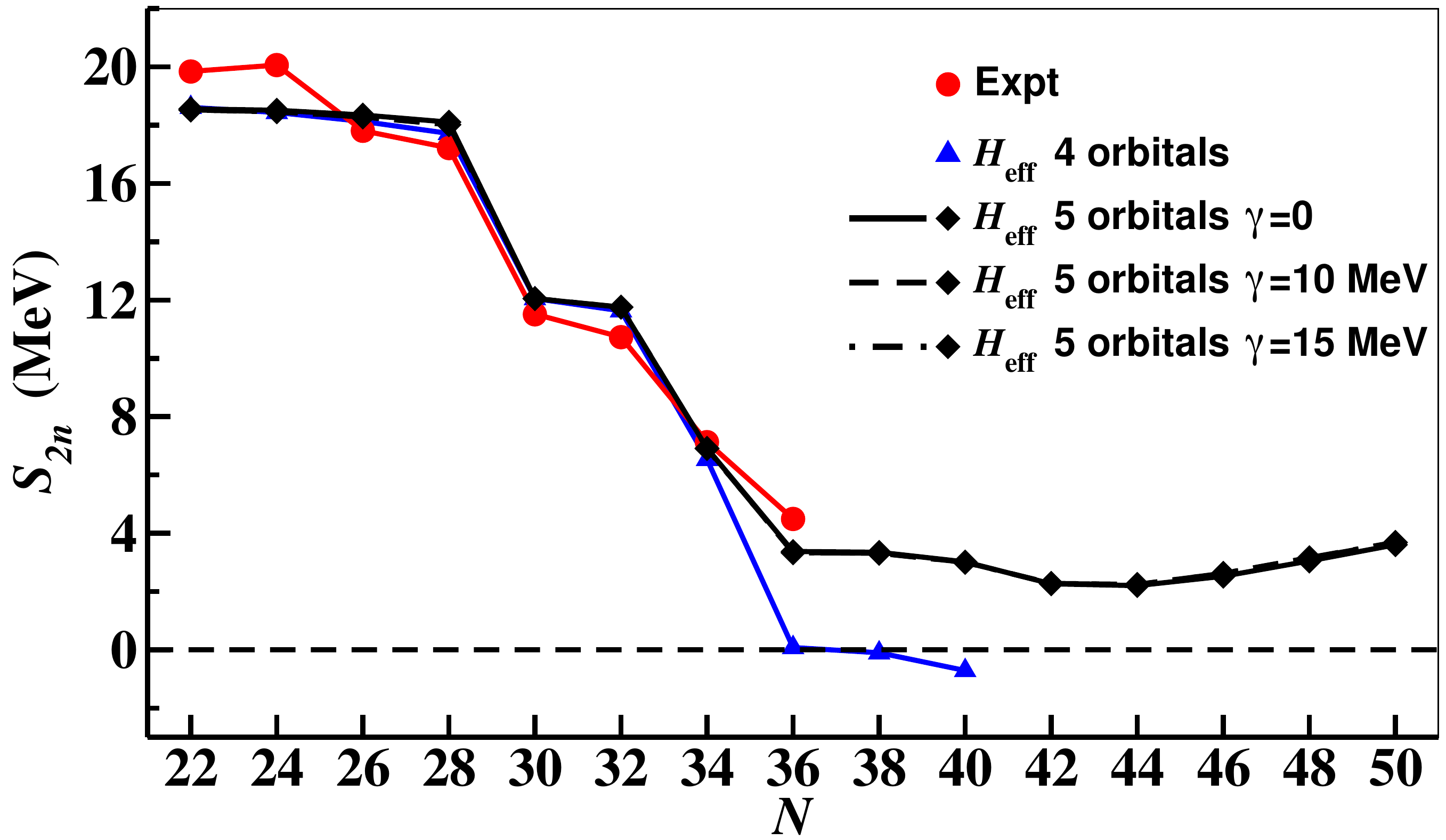}
\caption{Experimental and theoretical two-neutron separation energies
  for calcium isotopes from $N = 22$ to 50. Data are taken from
\cite{Audi03,Wienholtz13,Michimasa18}. See text for details.}
\label{S2nCa}
\end{center}
\end{figure}

The results of our calculations, performed by way of the shell-model
code KSHELL \cite{KSHELL}, are also presented in Fig. \ref{S2nCa}
(black diamonds, continuous line) to compare them with those we have
obtained in Ref. \cite{Ma19} where the model space we have employed
does not include the $0g_{9/2}$ orbital (blue triangles).
We report the experimental values as red dots and, as mentioned in
Section \ref{theory}, also the results obtained diagonalizing the
Hamiltonian $H'$ in Eq. \ref{hcom} with values of $\gamma=10$ MeV
(dashed black line) and 15 MeV (dash-dotted black line).

We note that closure properties, related to the filling of SP
orbitals, are reflected in the behavior of both experimental and
theoretical $S_{2n}$.

As can be seen, data and calculated values show a rather flat behavior
up to $N=28$, then a sudden drop occurs at $N=30$ 
that is a signature of the shell closure due to the $0f_{7/2}$ filling. 
Another decrease appears at $N=34$ because at that point the valence
neutrons start to occupy the $1p_{1/2}$ and $0f_{5/2}$ orbitals.
Then, from $N=36$ on, the calculated curve is rather flat matching the
filling of $0f_{5/2},0g_{9/2}$ orbitals.

The results obtained with both model spaces, the one considered in
present work with five neutron orbitals and the other with four orbitals
from Ref. \cite{Ma19}, follow closely the behavior of the experimental
$S_{2n}$ up to $N=34$, while those obtained in our previous work
provide an energy drop between $N=34$ and 36 much stronger than the
observed one.

It should be also pointed out that the results obtained with the
Hamiltonian $H'$ for values of the parameter $\gamma=10,15$ MeV
evidence that the spuriosities introduced by the center-of-mass motion are
under control for the calculation of two-neutron separation energies.

\begin{figure}[H]
\begin{center}
\includegraphics[scale=0.31,angle=0]{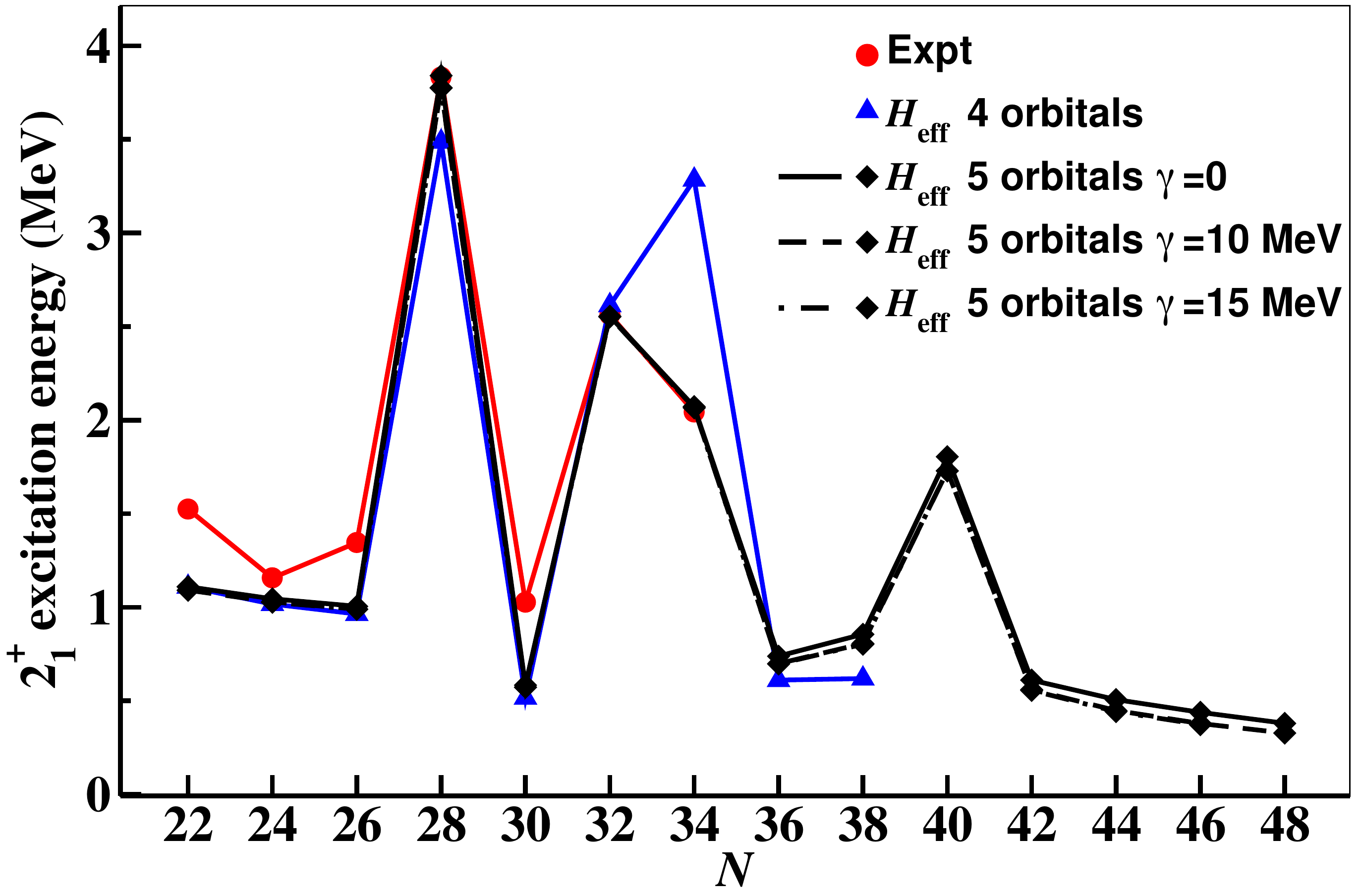}
\caption{Experimental and theoretical excitation energies of the yrast
  $J^{\pi}=2^+$ states for calcium isotopes from $N = 22$ to 50. See
  text for details.}
\label{J2pCa}
\end{center}
\end{figure}

This shell-closure properties of calcium isotopes can obviously be
also observed in the evolution of the excitation energies of the yrast
$J^{\pi}=2^+$ states with respect to the number of neutrons $N$, as
reported in Fig. \ref{J2pCa}.

It is noteworthy to observe that we obtain a better agreement with
experiment by including the $0g_{9/2}$ orbital.
In fact, within such a model space, we better reproduce the subshell
closure at $N=34$ and predict bound calcium isotopes at least up to
$N=50$.
Actually, the results obtained without $0g_{9/2}$ orbital provide, in
contrast with data, a raise of the excitation energy of the
$J^{\pi}=2^+_1$ state between $N=32$ and 34, and predict the calcium
drip line located at $N=38$ at variance with the recent observation of
a bound $^{60}$Ca \cite{Tarasov18}.

This testifies the need of a model space larger than the standard one
spanned by the $0f1p$ orbitals to perform a reliable investigation of
heaviest calcium isotopes.

In Fig. \ref{J2pCa} we have also reported the results we obtain by
employing the Hamiltonian $H'$ in Eq. \ref{hcom} with values of
$\gamma=10$ MeV (dashed black line) and 15 MeV (dash-dotted black
line).
As can be seen, the center-of-mass spuriosities provide a little
contribution also for the calculation of the excitation energies of
the yrast $J^{\pi}=2^+$.

It is now worth recalling that in the Introduction we mentioned about
the role that continuum states may play in isotopic chains approaching
their drip lines.
In a recent paper we investigated the neutron drip line of oxygen
isotopes \cite{Ma20}, which is experimentally placed at $N=24$, by
writing the many-body Hamiltonian in the Berggren basis and deriving a
\heff, built up in terms of the present chiral $NN$ and $NNN$
potentials,  that accounts for continuum states.
We have observed that this adds an additional repulsive effect to the
one provided by the $NNN$ component of the nuclear potential, and
leads to a better agreement with experiment too.

A similar procedure might be employed also to study calcium isotopes,
but the present limits of our computational resources prevent to
derive \heff~with the same accuracy we have done within the HO basis,
and we are working to implement continuum effects in the future also
for $fp$ isotopic chains.

There are two points that should be worth discussing in  connection
with the outcome of our calculations, namely the effects of many-body
correlations and the sensitivity of our results with respect to the
position of the neutron $0g_{9/2}$ orbital.

As regards the first one, we have already mentioned in the previous
section that we include the effect of second-order three-body diagrams,
which, for systems with more than 2 valence nucleons, account for the
interaction of the valence nucleons with core excitations as well as
with virtual intermediate nucleons scattered above the model space,
via the $NN$ potential \cite{Ellis77}.

This, as reported in Section \ref{theory}, has been done by deriving a
density-dependent two-body contribution at one-loop order from the
three-body correlation diagrams, and summing over the partially-filled
model-space orbitals, to overcome the limitations of our SM codes.
Here, we want to show the difference of the results of SM calculations
performed including or not the effects of such three-body correlations.

\begin{figure}[H]
\begin{center}
\includegraphics[scale=0.31,angle=0]{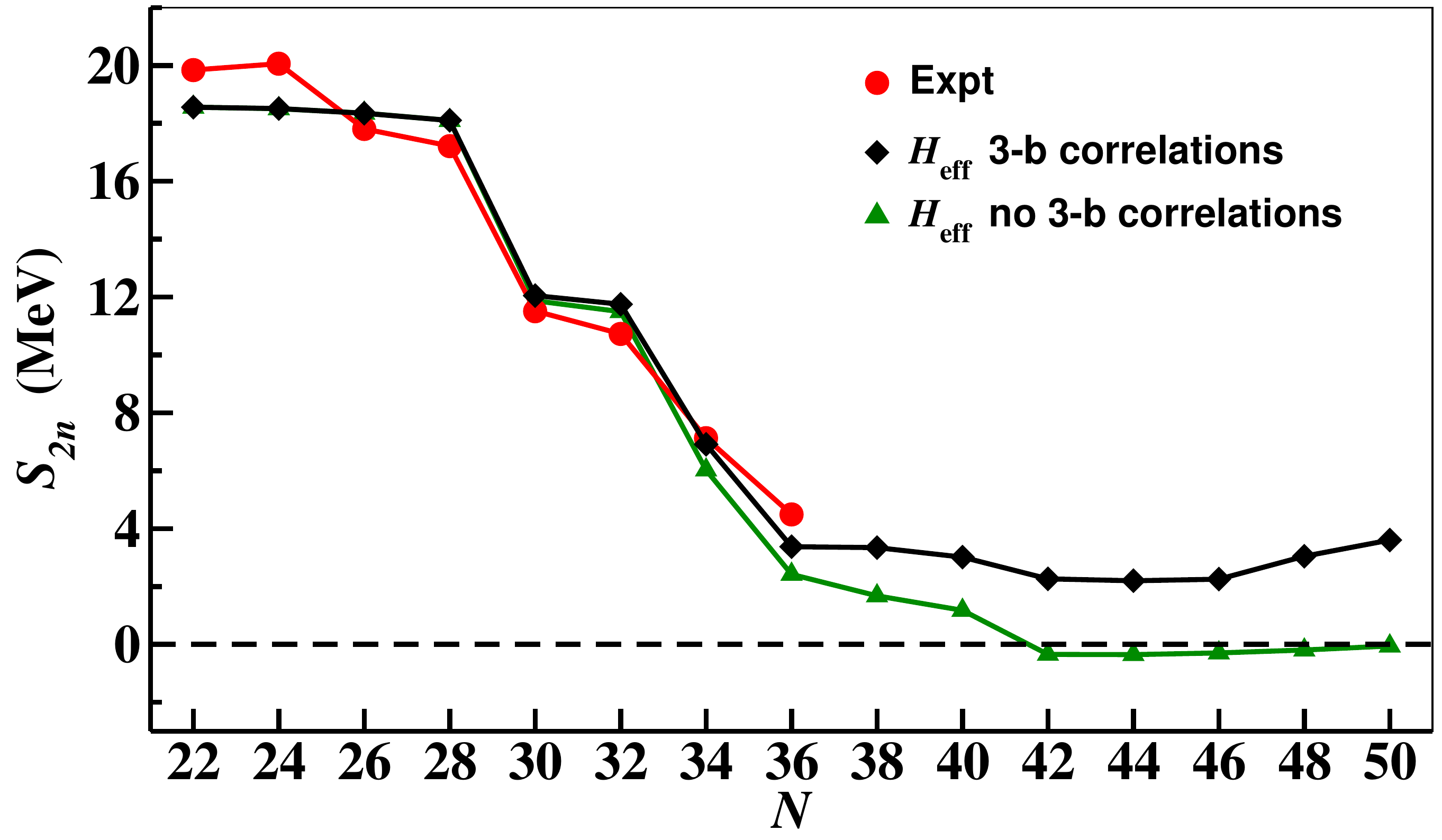}
\caption{Same as in Fig. \ref{S2nCa}, but including the results obtained
without three-body correlations (green triangles).}
\label{S2nCa_nocorr}
\end{center}
\end{figure}

\begin{figure}[H]
\begin{center}
\includegraphics[scale=0.31,angle=0]{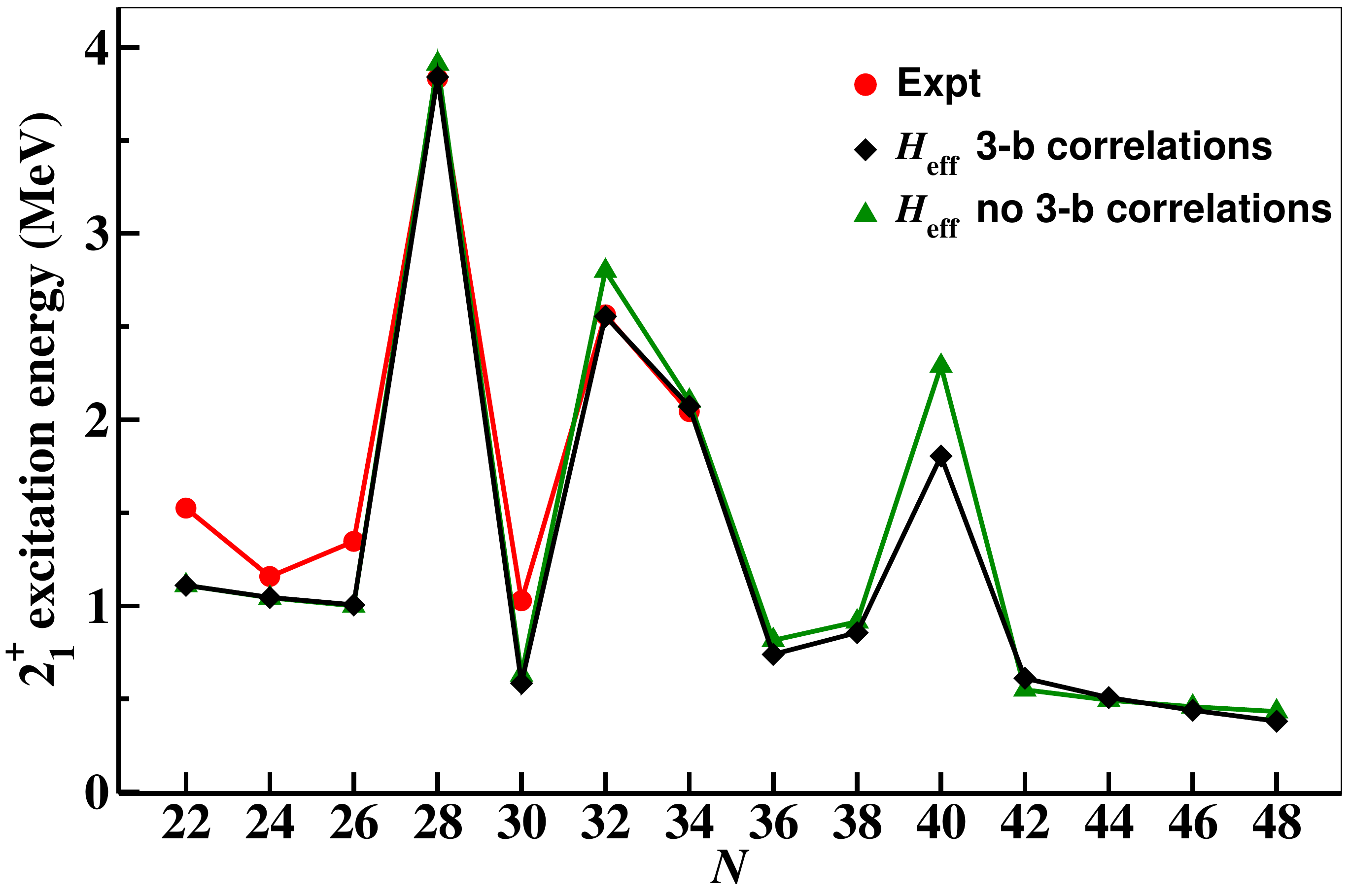}
\caption{Same as in Fig. \ref{J2pCa}, but including the results obtained
without three-body correlations (green triangles).}
\label{J2pCa_nocorr}
\end{center}
\end{figure}

To this end, in Figs. \ref{S2nCa_nocorr},\ref{J2pCa_nocorr} we compare
experimental two-neutron separation energies and excitation energies
of yrast $J^{\pi}=2^+$ states (red dots), respectively, of calcium
isotopes up to $N=50$, with the results of SM calculations obtained
including the effect of three-body correlations (black diamonds),
which means considering density-dependent \heffs, and with those
obtained with \heff~derived for just the two-valence nucleon system
(green triangles).

As can be seen, the effect of three-body correlations increases with
the number of valence nucleons, and starts to be substantial from
$N=34$ on.
The role of this effect is far more important for the ground-state
energies than the calculated excitation energies, and it is extremely
relevant to soundly determine the drip line.
In fact, without considering many-body correlations we see that the drip
line of calcium isotopes is located at $N=40$, while the attractive
contribution of second-order correlation diagrams shifts the last
bound nucleus at least to $^{70}$Ca.

The second point that should be examined is the correlation between
the SP spacings we have employed for our calculations, as reported in
Table \ref{tablespe}, and the evolution of the calculated two-neutron
separation energies.
Our SP spacings have been derived as the energy spectrum of the
effective Hamiltonian of one-valence nucleon systems, and they should
reproduce the experimental spectra of SP states in $^{41}$Ca (and
$^{41}$Sc for protons).

As a matter of fact, the experimental information about the
spectroscopic factors of $^{41}$Ca are rather scanty, and in fact the
observed counterpart of the second column in Table \ref{tablespe} is
missing.
However, there are three well-identified SP states in $^{49}$Ca, whose
spectroscopic factors with respect to doubly-closed $^{48}$Ca have
been measured \cite{ensdf,Uozumi94}, and they are reported together
with the calculated values in Table \ref{49Ca}.

\begin{table}[H]
\caption{Experimental negative-parity energy levels (in MeV) of
  $^{49}$Ca \cite{ensdf} $J^{\pi}=\frac{3}{2}^{-}_1,\frac{1}{2}^{-}_1
  \frac{5}{2}^{-}_{1,2}$ states, compared with the calculated
  ones. The values in parenthesis are the one-neutron pickup
  spectroscopic factors \cite{Uozumi94}.}
\begin{ruledtabular}
\begin{tabular}{ccc}
$J^{\pi}$ & Expt. & Calc. \\
\colrule
$\frac{3}{2}^-$  & 0.000 (0.84) & 0.000 (0.96) \\
$\frac{1}{2}^-$  & 2.023 (0.91) & 1.824 (0.97) \\
$\frac{5}{2}^-$  & 3.585 (0.11) & 3.710 (0.0001) \\
$\frac{5}{2}^-$  & 3.991 (0.84) & 4.191 (0.96) \\
\end{tabular}
\end{ruledtabular}
\label{49Ca}
\end{table}

The comparison between data and theory in Table \ref{49Ca}, as well as
the correct reproduction of the excitation energy of yrast $J^{\pi}=2^+_1$
state in $^{48}$Ca - which is strongly linked to the SP energy spacing
$\epsilon_{1p_{3/2}}-\epsilon_{0f_{7/2}}$ -, indicate that the
calculated SP spacings of natural-parity $0f1p$ orbitals are reliable.
As regards the position of the neutron $0g_{9/2}$ orbital, there is no
clear experimental indication if our calculated value is reasonable or
not.

We have then calculated the odd-even mass difference around $^{68}$Ni
that, because of the observed shell closure of nickel isotopes at
$N=40$, has a strong dependence on the energy gap between the
$0g_{9/2}$ and $0f_{5/2}$ orbitals, and whose experimental value is
3.2 MeV \cite{Audi03}.
The calculation cannot be performed exactly with our present
computational resources \cite{NATHAN,KSHELL}, so we truncate the
dimension of the eigenvalue problem by decomposing the eigenfunctions
in terms of broken pairs, and retaining only the components with
generalized seniority $v_g \leq 4$,  the results changing about
$0.5\%$ between $v_g=3$ and 4.
It turns out that our value is 4.2 MeV, 1 MeV larger than the
observed one, and indicating that the SP spacing
$\epsilon_{0g_{9/2}}-\epsilon_{0f_{5/2}}$ could be overestimated by
the same amount.

On the above grounds, we have investigated the sensitivity of our
results to the position of the $0g_{9/2}$ SP state, by raising and
lowering its energy by 1 MeV with respect to the calculated value.

\begin{figure}[H]
\begin{center}
\includegraphics[scale=0.31,angle=0]{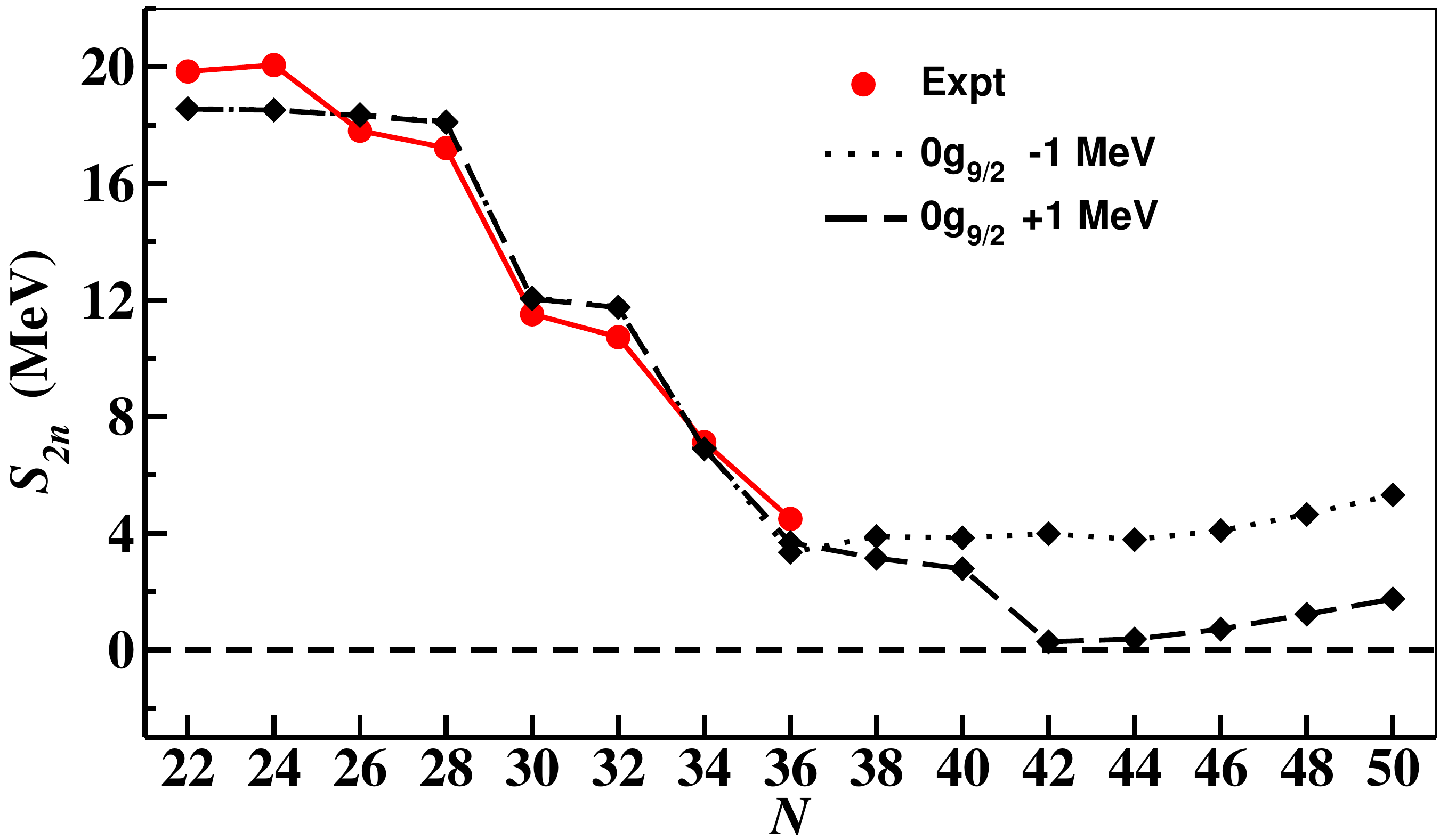}
\caption{Same as in Fig. \ref{S2nCa}, but with calculated values are
  obtained raising (dashed line) and lowering (dotted line)
  $\epsilon_{0g_{9/2}}$ by 1 MeV.}
\label{S2nCa_spread}
\end{center}
\end{figure}

\begin{figure}[H]
\begin{center}
\includegraphics[scale=0.31,angle=0]{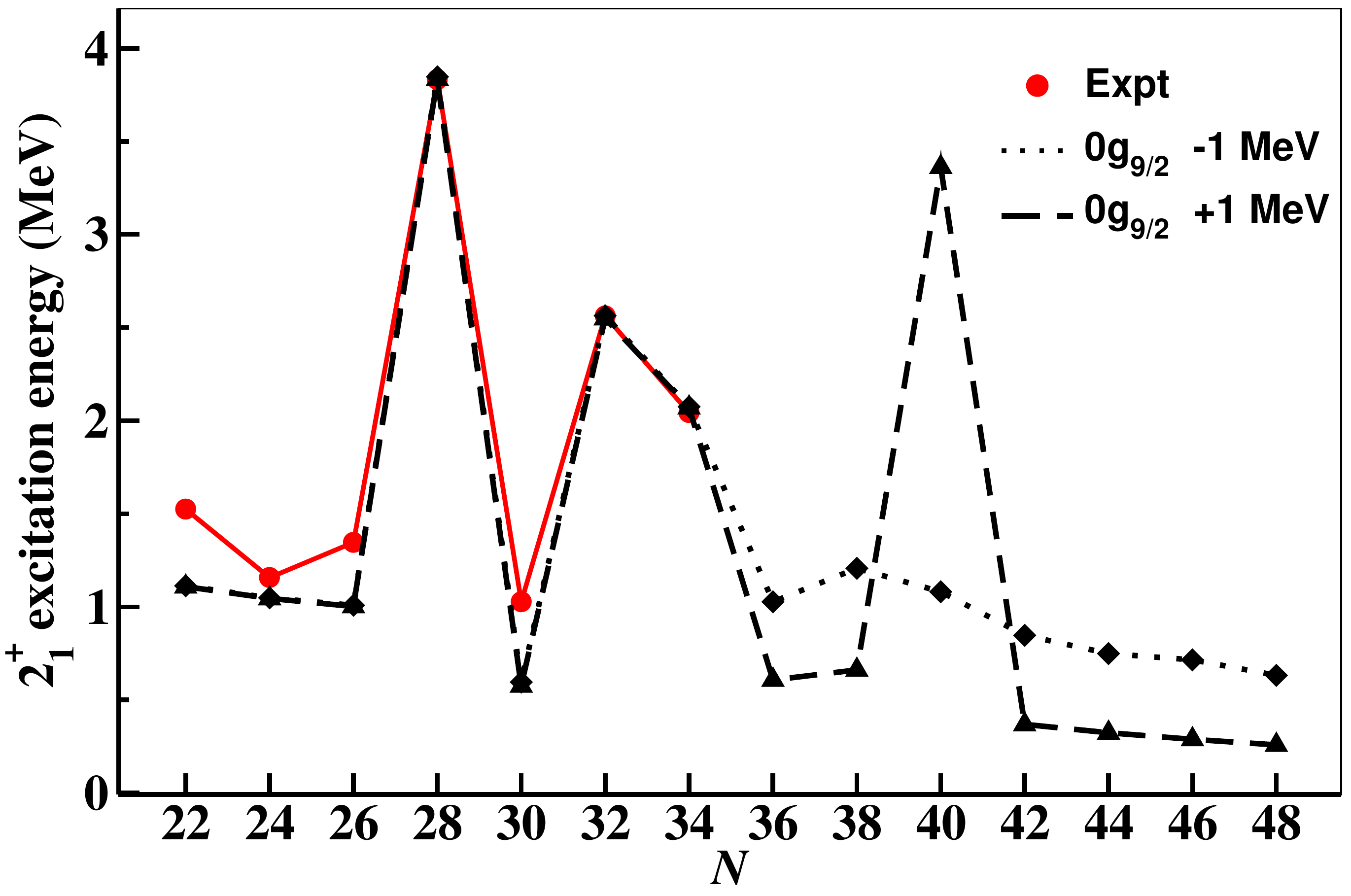}
\caption{Same as in Fig. \ref{J2pCa}, but with calculated values are
  obtained raising (dashed line) and lowering (dotted line)
  $\epsilon_{0g_{9/2}}$ by 1 MeV.}
\label{J2pCa_spread}
\end{center}
\end{figure}

In Figs. \ref{S2nCa_spread} and \ref{J2pCa_spread} we report the
results obtained with these two new values of $\epsilon_{0g_{9/2}}$
for both the two-neutron separation energies and the excitation
energies of yrast $J^{\pi}=2^+$, respectively.

We observe that up to $N=40$ the results barely differ each other as
well as from those with the set of SP energies in Table
\ref{tablespe}, and consequently the available experimental values
cannot discriminate about the choice of $0g_{9/2}$ SP energy.
From $N=40$ a clear distinction is observed: raising
$\epsilon_{0g_{9/2}}$ by 1 MeV we obtain that $^{62}$Ca is loosely
bound with respect to $^{60}$Ca, and a strong shell closure appears at
$N=40$.
This shell closure disappears lowering $\epsilon_{0g_{9/2}}$ by 1 MeV,
leading to the conclusion that a future measurement of the
experimental excitation energy of the yrast $J^{\pi}=2^+$ state may
provide insight on the location  of calcium isotopes drip line.

\section{Conclusions}
\label{endings}
This study is focussed on the location of the neutron drip line of
calcium isotopes, as predicted by nuclear shell-model calculations.

It is grounded on the results we have obtained in Ref. \cite{Ma19},
where, starting from two- and three-nucleon potentials derived within
the chiral perturbation theory, we have calculated effective SM
Hamiltonians able to reproduce the observed closure properties of
calcium isotopes and other $0f1p$ isotopic chains.
Here, in order to improve the depiction  of heavier systems, the
model space has been enlarged by adding the neutron $0g_{9/2}$
orbital.

Choosing a larger model space, we have improved the description of
the few observables in $^{54,56}$Ca with respect to the results obtained
without  $0g_{9/2}$ orbital, and have also been able to describe
$^{60}$Ca as a bound system, consistently with a recent experiment
\cite{Tarasov18}.

The main outcome of our investigation is, however, the fact that
according to our results the calcium isotopic chain is bound up to
$^{70}$Ca, at least, a result that is consistent with the recent
Bayesian analysis of different density-functional calculations
developed by Neufcourt and coworkers \cite{Neufcourt19}.
We have therefore also studied the relationship between our
theoretical tools and our prediction of the limits of calcium isotopes
as bound systems.

More precisely:
\begin{enumerate}
\item[a)] we have studied the role played by the inclusion of
  three-body correlations to calculate binding energies. As a matter
  of fact, calcium drip line would be placed at $N=40$ by neglecting
  these attractive contributions.
\item[b)] Since the position of the $0g_{9/2}$ single-particle energy
  is quite relevant when the filling of this orbital starts from
  $N=40$ on, we have investigated the sensitivity of both the two-neutron
  separation energies and the behavior of yrast $J^{\pi}=2^+$ excitation
  energy to this shell-model parameter.
  At present, available data do not allow to verify the reliability of
  our calculated prediction of $0g_{9/2}$ SP energy, at variance with
  the $0f1p$ orbitals, and we have found that the location of calcium drip
  line may be correlated with this quantity.
  In particular, our calculations indicate that the measurement of the
  yrast $J^{\pi}=2^+$ excitation energy in $^{60}$Ca could be pivotal
  to rule out predictions about the last bound calcium isotope.
\end{enumerate}

We consider this last point quite intriguing, since experimental
investigations of isotopes heavier than $^{60}$Ca may be very
challenging in a near future, and theory is in charge to point to
spectroscopic properties that at the same time should be easier to be
measured and provide informations about the ``terra incognita'' of
exotic nuclear systems.

\section*{Acknowledgements}
This work has been supported by he National Key R\&D Program of China
under Grant No. 2018YFA0404401, the National Natural Science
Foundation of China under Grants No. 11835001 and No. 11921006, and
the CUSTIPEN (China-US Theory Institute for Physics with Exotic
Nuclei) funded by the US Department of Energy, Office of science under
Grant No. DE-SC0009971.
We acknowledge the CINECA award under the ISCRA initiative through the
INFN-CINECA agreement, for the availability of high performance
computing resources and support, and the High-performance Computing
Platform of Peking University for providing computational resources.
G. De Gregorio acknowledges the support by the funding program
``VALERE'' of Universit\`a degli Studi della Campania ``Luigi
Vanvitelli''.

\bibliographystyle{apsrev}
\bibliography{biblio.bib}

\end{document}